\documentclass[12pt]
  {aipproc}

\usepackage{epsfig}  
\usepackage{amsmath}  
\usepackage{latexsym}  

\layoutstyle{8x11single}

\SetInternalRegister\hbadness{8000} % pseudo latin isn't breaking very well :-)

\newcommand\doingARLO[2][]{%
  \ifx\mmref\undefined #1\else #2\fi
}

\begin{document}

\title 
      {Towards a novel wave-extraction method for numerical relativity}


\author{A. Nerozzi}{
  address={Center for Relativity, University of Texas at Austin,
           Austin, Texas 78712, USA},
  email={anerozzi@physics.utexas.edu}
}

\iftrue
\author{M. Bruni}{
  address={Institute of Cosmology and Gravitation, University of Portsmouth,
           Portsmouth PO1 2EG, UK\\
           Dipartimento di Fisica, Universit\`a di Roma ``Tor
             Vergata'', via della Ricerca Scientifica 1, 00133 Roma, Italy},
  email={marco.bruni@port.ac.uk}
}

\author{L. M. Burko}{
  address={Department of physics, University of Alabama in Huntsville,
           Huntsville, Alabama 35899, USA},
  email={burko@email.uah.edu}
}

\author{V. Re}{
  address={School of Physics and Astronomy, University of Birmingham,
           Edgbaston, Birmingham, B15 2TT, UK},
  email={re@star.sr.bham.ac.uk}
}
\fi

% \copyrightholder{Acoustical Scociety of America}
\copyrightyear  {2006}

\begin{abstract}
We present the recent results of a research project aimed at constructing
a robust wave extraction technique for numerical relativity.
Our procedure makes use of Weyl scalars to achieve wave extraction.
It is well known that, with a correct choice of null tetrad, Weyl
scalars are directly associated to physical properties of the space-time
under analysis in some well understood way. 
In particular it is possible to associate $\Psi_4$ with
the outgoing gravitational radiation degrees of freedom, thus making it
a promising tool for numerical wave--extraction. The right choice of the
tetrad is, however, the problem to be addressed. We have made progress
towards identifying 
a general procedure for choosing this tetrad, by looking at transverse
tetrads where $\Psi_1=\Psi_3=0$.

As a direct application of these concepts, we present a numerical study 
of the evolution of a non-linearly disturbed black hole described 
by the Bondi--Sachs metric. This particular scenario allows
us to compare the results coming from Weyl scalars with the results
coming from the news function which, in this particular case, is
directly associated with the radiative degrees of freedom.
We show that, if we did not take particular care in choosing the
right tetrad, we would end up with incorrect results.
\end{abstract}

\date{\today}

\maketitle

\section{Introduction}

With the rising interest in gravitational--wave detection, the importance of numerical simulations
aimed at modelling possible sources of gravitational waves is growing. Possible sources of gravitational
waves include
binary systems of merging black holes, spiralling systems of two neutron stars or coalescing black hole--neutron star binaries. In order to simulate such physical events numerical relativity is required to study the strong
field region of the coalescence.
Most of the simulations which have been performed so far use formulation of Einstein's equations
such as ADM \cite{Arnowitt62, Lehner01a, Baumgarte:2002jm} or BSSN \cite{Baumgarte99,Shibata95}.
Using these formulations the problem of wave extraction arises, as none of the
variables used for the numerical evolution are related directly to the radiative degrees of freedom.

In this paper we review the research progress we have done in the field of wave
extraction using the Newman-Penrose formalism and, in particular, Weyl scalars
\cite{Beetle04,Nerozzi04, Burko04, Nerozzi06}. The final aim
of this project is to have at hand a quantity which can be computed directly from the numerically
evolved variables, and which is directly associated with the radiative degrees of freedom. The
strength of this technique is that it is completely general, i.e. it does not depend upon the
specific properties of the space-time under consideration. We apply these concepts to a specific
numerical example, namely to the characteristic initial value problem 
\cite{Bondi62, Sachs62}, in which we show the validity of our approach, and the problems that could
arise if we were not careful enough in computing Weyl scalars.

\section{Weyl Scalars}

When dealing with vacuum space-times, all the information about the curvature 
is encoded in the Weyl tensor $C_{abcd}$, which has ten independent components. All the ten degrees of freedom can be rewritten in a coordinate independent
way, through the introduction of the five complex Weyl scalars, defined as
\begin{eqnarray}
\Psi_0 &=& C_{pqrs}\ell^pm^q\ell^rm^s, \label{weyl0} \\
  \Psi_1 &=& C_{pqrs}\ell^pn^q\ell^rm^s, \label{weyl1} \\
  \Psi_2 &=& C_{pqrs}\ell^pm^q\bar{m}^rn^s, \label{weyl2}\\
  \Psi_3 &=& C_{pqrs}\ell^pn^q\bar{m}^rn^s, \label{weyl3} \\
  \Psi_4 &=& C_{pqrs}\bar{m}^pn^q\bar{m}^rn^s, \label{weyl4}
\end{eqnarray}
where $\ell^p$, $n^p$, $m^p$ and $\bar{m}^p$ are a set of four null
vectors, two real and two complex conjugates, satisfying the relations
$\ell^p n_p=-1$ and $m^p\bar{m}_p=1$. Of course the values of the scalars
will depend on the particular tetrad choice, so it is our aim to
identify the best tetrad choice for numerical relativity, so that
Weyl scalars are directly associated with the relevant physics degrees
of freedom \cite{Newman62a,Sachs62,Szekeres65,Teukolsky73}. 
One approach is to calculate Weyl scalars in a fiducial
tetrad, i.e. the easiest we can think of, and then perform tetrad
transformations in order to get to the physically relevant tetrad.
There are three sets of tetrad transformations which reflect, in tetrad
language, the Lorentz group of degrees of freedom for the basis vectors;
they are normally referred to as type I, type II and type III rotations;
in particular type III rotations are also known as spin/boost 
transformations (for more details see 
\cite{Chandrasekhar83,Nerozzi04}). What we need is a general criterion,
independent of the background space-time under study, for determining
what is the best tetrad transformation to perform in order to get 
to the physically relevant tetrad. As we will see in the next section,
this criterion is given by the notion of transverse frames.

\section{Transverse Frames}

We will show in this section and in the following that transverse frames 
are what we believe to be an attractive choice for performing wave extraction.
Following \cite{Beetle02, Beetle04, Nerozzi04}, we define a
transverse frame as a frame where $\Psi_1=\Psi_3=0$. Again following
the definitions given in those papers we call it frame instead of tetrad
because it actually identifies a family of tetrads connected by a type
III (spin/boost) transformation: the property $\Psi_1=\Psi_3=0$ is
in fact type III rotation invariant.

Finding a transverse frame is something that we can always achieve for
a generic space-time. In fact, it is clearly shown in 
\cite{Beetle02, Nerozzi04} that there are three transverse
frames for a generic Petrov Type I space-time. As a consequence
of this, the definition of a transverse frame is general and does
not depend at all on the properties of the background space-time
we are dealing with. 

Why transverse frames provide a general criterion for finding the
right tetrad is explained in the next section. Here we just want
to stress the attention on the generality of transverse frames: they
always exist for a general space-time and can always be found
using two tetrad rotations, namely a type I and a type II rotation
(see \cite{Nerozzi04} for further details).
                                                                                
\section{The Quasi-Kinnersley Frame}
\label{sec:kinframe}

The relevant property of a frame choice is that it has to converge 
to the Kinnersley frame when the space-time approaches a type D, i.e.
that of an unperturbed black hole. If this is true, then the
results found by Teukolsky (\cite{Teukolsky73}) assure us that
the Weyl scalars, in the linearized regime, are directly associated
to the physical properties of the space-time, and in particular $\Psi_4$
is associated to the outgoing gravitational radiation degrees of freedom.
To be more precise, in the linearized regime $\Psi_4$ can be directly
related to the TT gauge expression of the perturbed metric:

\begin{equation}
\Psi_4 = \frac{\partial^2 h^{TT}_{+}}{\partial t^2}+
i \frac{\partial^2 h^{TT}_{\times}}{\partial t^2}.
\label{eqn:psi4linearized}
\end{equation}

A \textsl{Kinnersley frame} for a type D space-time is a frame where
the two real tetrad null vectors coincide with the two repeated
principal null directions of the Weyl tensor. The fundamental 
property of a Kinnersley frame is that it has all Weyl scalars vanishing
except $\Psi_2$. This makes it an ideal frame where to build 
perturbation theory, and in fact this is what was done by Teukolsky
\cite{Teukolsky73} for a perturbed Kerr space-time.
                                                                                
In his original article, Kinnersley \cite{Kinnersley69}
makes a second step with an additional condition
that sets the spin coefficient $\epsilon$
to zero. This corresponds to fixing the additional degrees of freedom
coming from a spin/boost transformation, i.e. to identifying a particular
tetrad out of the Kinnersley frame. 

The quasi-Kinnersley frame is then defined as a general frame that converges
to the Kinnersley frame when the space-time approaches a type D.
We have shown in \cite{Nerozzi04} that one of the three transverse
frames is a quasi-Kinnersley frame; this comes as a direct consequence
of how a transverse frame sees the principal null directions of the
space-time under study. Finding the transverse frames, and especially
the one which is actually a quasi-Kinnersley frame, is for that reason
a general criterion to compute Weyl scalars in the right tetrad, in a
way that is completely independent from the properties of the
background metric.

In our study of the quasi-Kinnersley
frame we are also left with the problem of fixing the spin/boost
degrees of freedom. It is unlikely that the best choice is to
impose the condition $\epsilon=0$ in the quasi-Kinnersley frame,
as this condition does not fix univoquely the spin/boost degeneracy
of the frame, but only up to integration constants, so work is still
in progress to find an equivalent condition which is also easier
to implement.

\section{A numerical application: the Bondi problem}
\label{sec:bondi}

We have applied the concepts of quasi-Kinnersly frame to a
specific numerical simulation based on the characteristic initial
value problem, originally introduced by Bondi and Sachs 
\cite{Bondi62,Sachs62}. This approach has been extensively studied
in numerical relativity, for spherically symmetric systems
\cite{Corkill83,Gomez92a,Gomez94b,Clarke94,Clarke94b,Gomez96,Burko97},
axisymmetric systems \cite{Isaacson83,Gomez94a,Papadopoulos02} and 3D
systems \cite{Bishop97a,Bishop99,Gomez97a,Gomez97b}. The results
presented here are based on the procedure presented in \cite{Papadopoulos02},
where Bondi coordinates are used in axisymmetry to study a non-linearly
perturbed black hole.

In this particular case
the criterion to find the quasi-Kinnersley frame does not 
follow from the introduction of transverse frames, but from
simple considerations on the asymptotic behavior of the metric;
it turns out, in fact, that the tetrad we choose here as
quasi-Kinnerlsey tetrad, is not even transverse.
The metric in Bondi coordinates has the following expression:

\begin{equation}\label{eqn:bondimetric}
ds^2=-\left[\left(1-2\frac{M}{r}\right)e^{2\beta}-U^2r^2e^{2\gamma}\right]
\,dv^2+2e^{2\beta}dvdr
- 2Ur^2e^{2\gamma}dvd\theta +
r^2(e^{2\gamma}d\theta^2+e^{-2\gamma}\sin\theta^2d\phi^2), 
\end{equation}
where $M,U,\beta,\gamma$ are unknown functions of the coordinates $\left(v,r,
\theta\right)$.
Within this framework, the
Einstein equations decompose into three hypersurface equations and one
evolution equation, as given below in symbolic notation

\begin{subequations}
\label{eqn:bondiev}
\begin{eqnarray}
\Box^{\left(2\right)}\psi &=& \mathcal{H}_{\gamma}\left(M,\beta,U,\gamma\right), \label{eqn:gammaeqn} \\
\beta_{,r} &=& \mathcal{H}_{\beta} \left(\gamma\right), \label{eqn:betaeqn} \\
U_{,rr} &=& \mathcal{H}_U \left(\beta, \gamma\right), \label{eqn:ueqn} \\
M_{,r} &=& \mathcal{H}_M \left(U, \beta, \gamma\right), \label{eqn:meqn}
\end{eqnarray}
\end{subequations}
where $\Box^{\left(2\right)}$ is a 2-dimensional wave operator, $\psi=r\gamma$,
and the various $H$ symbols are functions of the Bondi variables.
A more detailed description of the system defined in Eq.~(\ref{eqn:bondiev})
can be found in \cite{Gomez94a,Isaacson83}. The algorithms for integrating
that system of equations numerically are well known and tested. 

The next step is to decide the initial conditions for our numerical evolution:
the metric introduced in Eq.~(\ref{eqn:bondimetric})
describes a static Schwarzschild black hole if
we set, in the Bondi frame,  all the functions except $M$
to zero everywhere in the domain.
$M$ is chosen to be the Schwarzschild mass $M_0$ of the black hole.
Besides, outgoing gravitational radiation as a perturbation is introduced
by the function $\gamma$: in fact $\gamma$ is a spin-2 field
and is actually related to the radiative degree of freedom. Choosing
an initial shape for $\gamma$ means in practice choosing the initial profile of
outgoing gravitational waves. The typical profile that we set up
for initial data has the following expression

\begin{equation}
\gamma\left(r,\theta\right) = \frac{\lambda}{\sqrt{2\pi}\sigma}e^{-\frac{\left(r-r_c\right)^2}{\sigma^2}} Y_{2lm}\left(\theta\right),
\label{eqn:gammapert}
\end{equation}
where $\lambda$ is the amplitude of the perturbation, $r_c$ is the center
of the perturbation in r and $\sigma$ its variance; finally $Y_{lm}$ is the
spherical harmonic of spin 2. For the numerical simulations we will consider
here, we will set the initial data to be a pure quadrupole wave, i.e. $l=2$ and
$m=0$ in $Y_{2lm}\left(\theta\right)$. 

\subsection{Tetrad Choices}
\label{sec:tetradchoice}

Choosing the quasi--Kinnersley tetrad for our specific case is quite
simple: the equations for the Bondi functions do not determine those
completely, in fact we have of course the freedom coming from the
integration constants. If we set those integration constants to be zero
(Bondi frame) we know what is the asymptotic behavior of all the
functions when we are far from the source, so 
$\gamma$, $U$ and $\beta$ tend to zero, while $M$ tends to the
Schwarzschild mass $M_0$ of the black hole.

In the Schwarzschild limit the Kinnersley tetrad in our coordinate system
is given by

\begin{subequations}
\label{eqn:kinbondi}
\begin{eqnarray}
\ell^{\mu}&=&\left[\frac{2r}{r-2M},1,0,0\right],
  \label{eqn:schwarztetradnulll} \\
  n^{\mu}&=&\left[0,-\frac{r-2M}{2r},0,0\right],
  \label{eqn:schwarztetradnulln} \\
  m^{\mu}&=&\left[0,0,\frac{1}{\sqrt{2}r},\frac{i}
{\sqrt{2}r\sin\theta}\right].
\label{eqn:schwarztetradnullm}
\end{eqnarray}
\end{subequations}
From this we get that a general tetrad that converges to the
Kinnersley tetrad, using our knowledge of the asymptotic 
behavior of the functions, is given by

\begin{subequations}
\label{eqn:quasikinbondi}
\begin{eqnarray}
\ell^{\mu}&=&\left[\frac{2}{\left[\left(1-2M/r\right)e^{4\beta}-
  U^2r^2e^{2(\gamma+\beta)}\right]},e^{-4\beta},0,0\right], \nonumber \\
  \label{eqn:bonditetradnulll} \\
  n^{\mu}&=&\left[0,-\frac{\left[\left(1-2M/r\right)e^{2\beta}-
    U^2r^2e^{2\gamma}\right]}{2},0,0\right],
    \label{eqn:bonditetradnulln} \\
    m^{\mu}&=&\left[0,\frac{rUe^{(\gamma-2\beta)}}
{\sqrt{2}},\frac{1}{\sqrt{2}re^{\gamma}},\frac{i}
{\sqrt{2}r\sin\theta e^{-\gamma}}\right].
\label{eqn:bonditetradnullm}
\end{eqnarray}
\end{subequations}

Another way to pick the tetrad for computing Weyl scalars is
use algebraic manipulation packages like Maple and GRTensor. For the
Bondi metric they would come out with the following expression
for the tetrad:

\begin{subequations}
\label{eqn:nonquasikinbondi}
\begin{eqnarray}
\ell^{\mu}&=&\left[
0,-e^{-4\beta},0,0\right], \nonumber \\
\label{eqn:bonditetradnulll2} \\
n^{\mu}&=&\left[e^{2\beta},\frac{\left[\left(1-2M/r\right)e^{2\beta}-
U^2r^2e^{2\gamma}\right]}{2},0,0\right], \nonumber \\
\label{eqn:bonditetradnulln2} \\
m^{\mu}&=&\left[0,
\frac{rUe^{(\gamma-2\beta)}}{\sqrt{2}},
\frac{1}{\sqrt{2}re^{\gamma}},
\frac{i}{\sqrt{2}r\sin\theta e^{-\gamma}}\right].
\label{eqn:bonditetradnullm2}
\end{eqnarray}
\end{subequations}
GRTensor gives this expression because it constructs the tetrad 
starting from the $\ell^{\mu}$ vector, which is assumed to be lying
on the null foliation, leading to the expression $\ell_{\mu}=\delta_{\mu 0}$.
The contravariant components are then given by
Eq.~(\ref{eqn:bonditetradnulll2}). Once $\ell^{\mu}$ is fixed, the other
tetrad vector expressions are found by imposing the normalization conditions
between the vectors in the Newman-Penrose formalism.

\section{Numerical Results}
\label{sec:results}

\begin{figure}
\includegraphics[height=8cm, width=15cm]{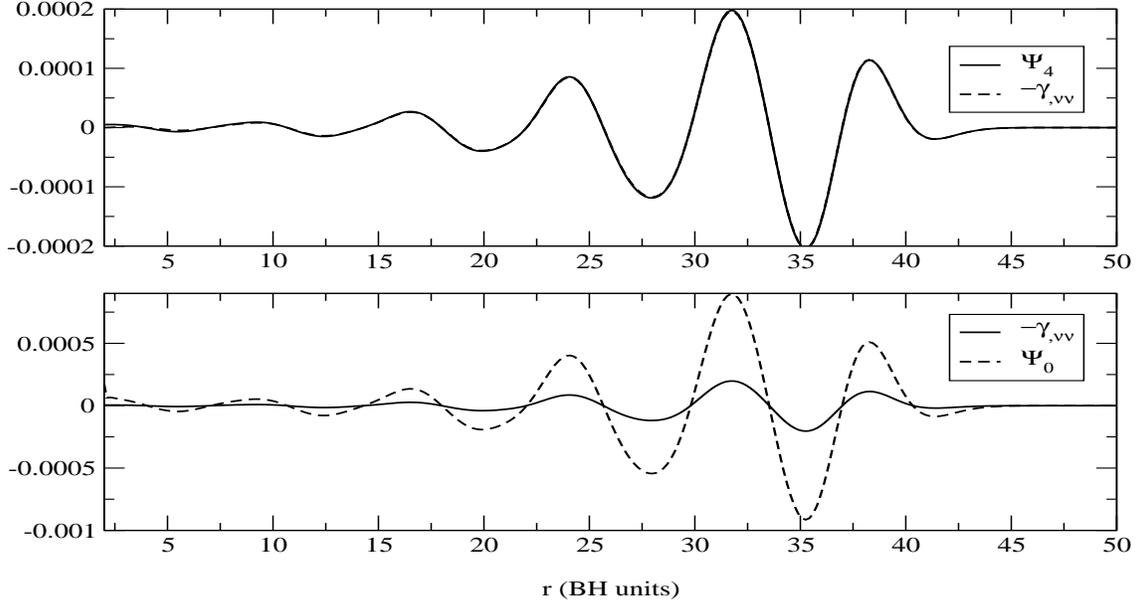}
\caption{Comparison of different news functions. The upper figure shows
the comparison of the second time derivative of the Bondi function $\gamma$
with the value of $\Psi_4$ at time $v=80$, i.e. in the linearized regime.
It is clear that the two functions are in very good agreement. The same
thing is done in the lower figure but this time using the second tetrad,
in this case the outgoing contribution is contained in $\Psi_0$ but it
is clear that the two news function do not agree, in fact we show in 
\cite{Nerozzi06} that this second tetrad turns out to be boosted with respect
to the quasi-Kinnersley tetrad.}
\label{fig:compare}
\end{figure}

We show here an example of the results for a numberical simulation
of the Bondi metric, with wave extraction performed by using both
news function and Weyl scalars. 
We set up an initial Schwarzschild black hole plus an
initial quadrupole perturbation on $\gamma$ using 
Eq.~(\ref{eqn:gammapert}). In this particular case we set 
$\lambda=0.1$, $r_0=3$ and $\sigma=1$. The Bondi functions are then
evolved and the wave is extracted using the news function, namely
the time derivative of the Bondi function $\gamma$. We try to achieve
the same results by using Weyl scalars. It is easy to see that, by
means of equation \ref{eqn:psi4linearized}, 
$\Psi_4$ has to be equal to the second time
derivative of $\gamma$ in the linearized regime provided it is 
computed in the right tetrad. So a very straightforward test to prove
that we are in the right tetrad is to compare $\Psi_4$ with 
$\gamma_{,vv}$ in the linear regime, i.e. at late times and far from
the source. 

We have done this for our simulation, and a significant result is 
presented in Fig.~(\ref{fig:compare}). Here the news function is 
compared with $\Psi_4$ for the first tetrad, and with $\Psi_0$ in the
second tetrad. The reason why we consider $\Psi_0$ for the second
tetrad is due to the fact that in the second tetrad the $\ell$ null vector is
ingoing instead of outgoing, and this changes the roles of the scalars
in such a way that the outgoing contribution is in $\Psi_0$ instead
of $\Psi_4$ (see \cite{Nerozzi06} for further details). 
It is clear that the value for $\Psi_0$ in the second tetrad does
not give the correct value for the gravitational radiation emitted
by the source. Speaking in terms of tetrad rotations, this means
that the second tetrad is actually boosted with respect to the 
quasi-Kinnersley tetrad. By comparing the two tetrads, we can compute
the coefficient of the boost transformation that relates them, which is
given by

\begin{equation}
A=\frac{2r}{r-2M}.
\label{eqn:typeIIIcoeff}
\end{equation}
This means that $\Psi_0$ in the second tetrad is related to $\Psi_4$
in the first tetrad by the relation

\begin{equation}
\left(\Psi_4\right)_{\mathcal{T}_1} =
\left(\frac{r-2M}{2r}\right)^2\left(\Psi_0\right)_{\mathcal{T}_2},
\label{eqn:finalrel}
\end{equation}
where $\mathcal{T}_1$ and $\mathcal{T}_2$ denote the two tetrads. It can
be shown \cite{Nerozzi06} that the results in the two tetrads coincide if we
consider this additional boost factor. The derivation of this 
boost coefficient has been possible thanks to the determination of the
quasi-Kinnersley tetrad: had we started directly from tetrad $\mathcal{T}_2$
we would have incorrectly associated the value of $\Psi_0$ to the gravitational
radiation.

\section{Conclusions}

We have presented the recent progress made in the field of wave extraction using
Weyl scalars. We believe that a promising technique for solving generally the
problem of the tetrad choice is to look at transverse frames, i.e. those
families of tetrads where $\Psi_1=\Psi_3=0$. This approach has the advantage
that for such families of tetrads it is possible to associate $\Psi_4$ with the
outgoing radiative degrees of freedom in a background independent way. Our 
project still needs further work concerning how to break the spin/boost
degeneracy in a background independent way.
We have applied these concepts to a specific numerical example, using the
Bondi-Sachs metric. In this particular case we can avoid the the spin/boost
degeneracy  problem and choose directly a quasi-Kinnersley tetrad for wave
extraction, that is the one associated to the news function. Knowing this
tetrad then allows us to explicitly show how choosing a different tetrad for
wave extraction leads to wrong results.

\doingARLO[\bibliographystyle{aipproc}]
          {\ifthenelse{\equal{\AIPcitestyleselect}{num}}
             {\bibliographystyle{arlonum}}
             {\bibliographystyle{arlobib}}
          }

\end{document}